# Infrared Polarization Imaging-based Non-destructive Thermography Inspection

Xianyu Wu, Bin Zhou, Peng Lin, Rongjin Cao, Feng Huang

***Abstract*—Infrared pulse thermography non-destructive testing (NDT) method is developed based on the difference in the infrared radiation intensity emitted by defective and non-defective areas of an object. However, when the radiation intensity of the defective target is similar to that of the non-defective area of the object, the detection results are poor. To address this issue, this study investigated the polarization characteristics of the infrared radiation of different materials. Simulation results showed that the degree of infrared polarization of the object surface changed regularly with changes in thermal environment radiation. An infrared polarization imaging-based NDT method was proposed and demonstrated using specimens with four different simulated defective areas, which were designed and fabricated using four different materials. The experimental results were consistent with the simulation results, thereby proving the effectiveness of the proposed method. Compared with the infrared-radiation-intensity-based NDT method, the proposed method improved the image detail presentation and detection accuracy.**

## I. Introduction

INFRARED imaging non-destructive testing (IR-NDT) methods are widely applied for the inspection of metal materials, fiber-reinforced polymer composite materials, and thermal barrier coatings owing to their fast, intuitive, and non-contact inspection of large areas [1]. They can be categorized as active and passive, depending on whether external thermal excitation is applied to form a thermal contrast between defective and non-defective areas in the testing process. To produce significant thermal contrast between defect and non-defect areas, active IR-NDT technology typically uses pulsed or modulated thermal excitation to heat specimens. The most common sources of heating include optical (halogen lamps [2],[3] and lasers [4]), mechanical (vibrational thermography [5]), and electromagnetic sources [6],[7]. In IRT methods, an infrared thermal imager is typically used to record changes in the surface temperature field of test specimens. Commonly used thermal imagers include cooled and uncooled infrared detectors (i.e., microbolometers). Compared with cooled infrared detectors, uncooled infrared detectors are economical, small, lightweight, and power-efficient, while meeting detection accuracy requirements; therefore, uncooled infrared detectors have broad application prospects in industry [8].

Microbolometers [9] are thermal imaging sensors produced using infrared radiation-absorbing materials. When the infrared radiation of an object irradiates the sensor, the electrical resistance of each sensor pixel changes, and the temperature field or infrared radiation information of the object surface is measured by recording the change in the electrical resistance value.

However, the aforementioned infrared imagers can only be used to record the infrared radiation intensity image of the object's surface; they cannot detect the polarization information of infrared radiation [10]. The infrared radiation polarization characteristics of an object are closely related to its surface and subsurface materials and are affected by its physical characteristics. Moreover, the polarization characteristics of an object can provide more useful information than the infrared radiation intensity. The polarization information can not only reflect the characteristics of object surfaces, microstructures and materials, but also has a wide range of application prospects in object identification, road detection, and semantic segmentation [11]. However, existing IR-NDT methods only use the intensity information of the target infrared radiation to detect subsurface defects in specimens.

A variety of polarization imaging techniques have been developed and applied in the research field. According to the differences in the methods of obtaining polarization state images at different angles, infrared polarization imaging techniques can be categorized into division-of-time [12], division-of-amplitude [13], division-of-aperture [14], and division-of-focal-plane polarization imaging methods [15],[16]. The division-of-time infrared polarization imaging system is inexpensive and has high measurement accuracy. However, it has a slow imaging speed and cannot be applied to acquire polarization images of dynamic scenes, which hinders its application in the field of NDT. The division-of-amplitude polarization imaging system acquires real-time multi-polarization state images by splitting light into different optical paths and synchronizing multiple infrared imaging sensors. Its optical-mechanical structure is complex and requires precise calibration procedures. The division-of-aperture polarization imaging system uses a lens array to focus objects on separate infrared imaging sensors. Each sensor acquires a different polarization state image. Therefore, the fields of view of the images are different, and post-processing operations are required for image registration and alignment. A division-of-focal-plane (DoFP) polarization-imaging detector uses a pixelized micropolarizer array to measure different polarization states. In addition, this detector can realize real-time polarization imaging of dynamic scenes and overcome the shortcomings of other polarization imaging systems. Moreover, this is the most suitable method for IR-NDT applications.

In recent years, theoretical and experimental studies have been conducted in the field of infrared polarization imaging [17],[18]. Several studies have shown that both the reflected and spontaneous radiations of a target contain polarization information. Moreover, the observation angle [19], surface roughness [20], and material and physical properties of the target [21],[22] have a significant impact on the polarization detection of the target. Liu et al.[23] deduced an analytical model for the infrared polarization degree of a target surface based on a polarization bidirectional reflectance distribution function. Through theoretical simulations and experiments, they concluded that environmental thermal radiation affects the regular change in the infrared polarization degree of the target. Xu et al. [24] proposed a method for detecting coating and substrate materials based on infrared spectral polarization measurements. Thus, the subsurface material characteristics of a target object can be detected using the infrared radiation polarization characteristics of the target object surface, making the NDT of subsurface defects possible using these characteristics.

In this study, a mathematical model for the infrared polarization radiation of an object surface was first established based on the infrared polarization characteristics of the object surface [23],[25]. The model was used to describe the influences of environmental radiation intensity changes, target surface materials, and substrate materials on the polarization characteristics of infrared radiation of the target surface. Subsequently, a simulation was conducted, and we proved that the NDT of subsurface damage can be conducted by analyzing the change in the infrared radiation polarization information of the target surface. Finally, we fabricated a carbon-fiber-reinforced polymer (CFRP) specimen embedded with four different types of artificial defects and verified the feasibility of the NDT method based on the polarization degree of the infrared radiation of the surface of the target object. Fast Fourier transform (FFT) [26] and principal component analysis (PCA) [27] processing were conducted for the captured infrared radiation intensity image sequence and polarization image sequence. The image processing results confirmed that the NDT method based on the polarization characteristics of infrared radiation can overcome the limitations of infrared radiation intensity-based NDT methods; this makes it difficult to distinguish different artificial defective materials with small temperature differences.

## II. METHOD

### A. *Polarization principle of infrared thermal radiation*

For the NDT of subsurface defects, inspection methods based on infrared radiation intensity or thermography intensity images have been developed based on the temperature difference between defective and non-defective areas. Therefore, when the temperature difference between these areas is small, the defect area cannot be distinguished from the inspection result images. In this context, for the first time, we used the difference in the degree of linear polarization (DoLP) between defect and non-defect areas for non-destructive testing, thereby overcoming the shortcomings of infrared thermography-based NDT methods. First, a model was established for the quantitative analysis of the polarization characteristics of the infrared radiation of the object's surface.

According to [23] and [25], the DoLP model of an object surface is given by (1).

$$DoLP = \frac{\frac{1}{8\pi\sigma_m^2}|I_E - I_{obj}|}{I_{obj} + \frac{I_E - I_{obj}}{8\pi\sigma_m^2}\int_0^{2\pi}\int_0^{\pi/2}\frac{1}{\cos^4\psi_N} \cdot \frac{\exp[-\tan^2\psi_N/(2\sigma_m^2)]}{\cos\psi_i}\sin\psi_r \cdot (R_S + R_P)d\psi_r d\phi_r} \cdot$$

$$\left\{\left\{\int_0^{2\pi}\int_0^{\pi/2}\frac{1}{\cos^4\psi_N} \cdot \frac{\exp[-\tan^2\psi_N/(2\sigma_m^2)]}{\cos\psi_i}\cos 2\eta_r \cdot \sin\psi_r \cdot (R_S - R_P)d\psi_r d\phi_r\right\}^2 + \left\{\int_0^{2\pi}\int_0^{\pi/2}\frac{1}{\cos^4\psi_N} \cdot \frac{\exp[-\tan^2\psi_N/(2\sigma_m^2)]}{\cos\psi_i}\sin 2\eta_r \cdot \sin\psi_r \cdot (R_P - R_S)d\psi_r d\phi_r\right\}^2\right\}$$ (1)

where $\psi$ and $\phi$ are the zenith and azimuth angles, respectively. The subscripts $i$ and $r$ indicate the quantities associated with the incident and reflected radiants, respectively. $I_{obj}$ is the emitted intensity from the isothermal blackbody. $I_E$ is the incident intensity from the surrounding environment. $\sigma_m$ is the surface roughness of the target material. $\psi_N$ is the angle between the normals of the object and microfacet surfaces. $R_S$ and $R_P$ represent the Fresnel reflectivities of the orthogonal and parallel components, respectively.

However, in general, if $\psi_N$ is small and only the incident and reflected lights of the same plane are considered, the thermal radiation is modeled as shown in Fig. 1, and the DoLP can be simplified using (2):

$$DoLP_{simplified} = \frac{|\alpha - 1| \times |R_S - R_P|}{8\sigma_m \cos\psi_i \pm (\alpha - 1) \times (R_S + R_P)} \quad (2)$$

$$R_S(\lambda, \psi_i) = \left|\frac{n_{air}(\lambda)\cos\psi_i - n_m(\lambda)\cos\psi_j}{n_{air}(\lambda)\cos\psi_i + n_m(\lambda)\cos\psi_j}\right|^2 \quad (3)$$

$$R_P(\lambda, \psi_i) = \left|\frac{n_m(\lambda)\cos\psi_i - n_{air}(\lambda)\cos\psi_j}{n_m(\lambda)\cos\psi_i + n_{air}(\lambda)\cos\psi_j}\right|^2 \quad (4)$$

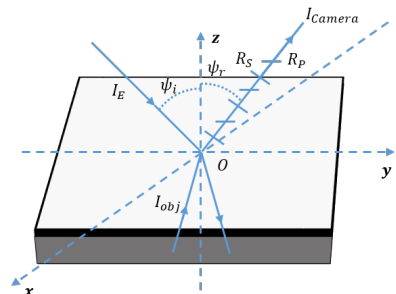

Fig. 1. Schematic of the simplified thermal radiation transfer model.

The ± sign in (2) depends on the ratio of the incident intensity from the surrounding $I_E$ to the emitted intensity from the isothermal blackbody $I_{obj}$ expressed as $\alpha = I_E/I_{obj}$. For $\alpha$ greater than 1, a positive sign is assumed, while a negative sign is assigned for $\alpha$ less than 1. $n_{air}$ and $n_m$ are the refractive indices of air and the target material, respectively. $\psi_i$ and $\psi_j$ are the incidence angle of environmental radiation and the refraction angle formed by the light passing through the target surface, respectively. Therefore, when the material properties and incident angle of environmental radiation are constant, the change in DoLP depends on the change in $\alpha$. By using the derivative of $\alpha$ from (2), we obtain

$$\frac{\partial DoLP_{simplified}}{\partial \alpha} = \frac{\pm 8\sigma_m \cos\psi_i \times |R_S - R_P|}{[8\sigma_m \cos\psi_i \pm (\alpha-1)\times(R_S + R_P)]^2} \quad (5)$$

Similarly, the positive and negative signs in (5) depend on $\alpha$. When $\alpha$ is greater than 1, the derivative obtained using (5) is positive; that is, the DoLP increases with an increase in $\alpha$. When $\alpha$ is less than 1, the derivative obtained using (5) is negative, and the DoLP decreases as $\alpha$ increases. Therefore, when the incident intensity from the surrounding environment is constant, the DoLP of the target object changes accordingly as the thermal radiation emitted by the target object changes continuously. Based on the aforementioned phenomena, an NDT method based on the polarization information of the infrared radiation of a target object can be established. After the test specimen was subjected to a short pulse of thermal excitation, the temperature change and the temperature change rate of these areas were inconsistent because the heat absorptions of the defective and non-defective areas were inconsistent,. Therefore, the values of $\alpha$ at different positions on the specimen are different, and changes in $\alpha$ in the defective area is faster or slower than that in the non-defective area, thereby affecting the speed change in the DoLP.

In this study, we compared the defect detection capability from a sequence of thermograms (henceforth called DoFP) and the related DoLP image sequence. The infrared polarized images captured at different observation angles were processed, and a raw DoFP intensity image sequence and an infrared radiation DoLP image sequence were obtained. To compensate for the pixel resolution loss after the DoLP calculation, upsampling interpolation on the four polarization state image sequences was performed [28]. To reduce the impact of the noise of the raw DoFP image on the DoLP image, we assumed that the mean and variance of the noise of each channel of the raw DoFP image $X_{.,j}$ for all the channels $j$, $j \in [0, 45, 90, 135]$ are 0 and $\sigma^2$, respectively. Then the noise of $S0$ is expressed as

$$n(S0) = [n(X_{.,0}) + n(X_{.,45}) + n(X_{.,90}) + n(X_{.,135})]/4$$

The mean and variance of $n(S0)$ are 0 and $\sigma^2/4$, respectively, implying that the standard deviation of the noise of $S0$ is $\sigma/2$, which is half of that for $X$. Therefore, we used $S0$ as the guidance image to perform guided filtering on the DoLP image [29]. The DoFP image processing pipeline is shown in Fig. 2.

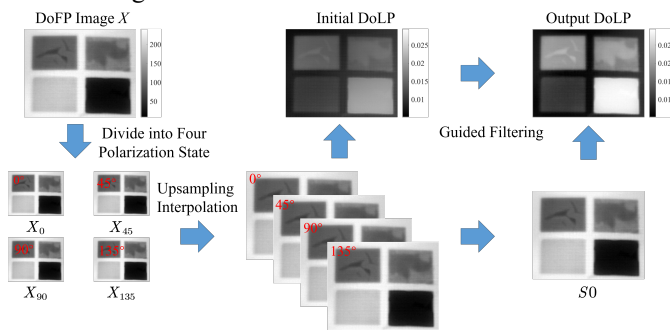

Fig. 2. Schematic of the proposed DoFP image processing pipeline.

### B. Image processing principle

Similar to an NDT method based on an infrared radiation intensity image sequence, NDT can also be performed using FFT or PCA processing for a polarization image sequence recording changes in the DoLP. Similarly, pulsed phase thermography uses a Fourier transform to convert time-domain data into frequency-domain data and then analyzes and compares the amplitude and phase information images for non-destructive testing.

$$F_{x,y}(j\Omega)|\Omega = k\Omega_0 = F(jk\Omega_0) =$$
$$\frac{1}{N}\sum_{n=0}^{N-1} T(n) e^{-\frac{2k\pi ni}{N}} = R(k) + jI(k) \quad (6)$$

$$|F(k)| = \sqrt{R^2(k) + I^2(k)}, \varnothing(k) = \arctan\frac{I(k)}{R(k)}, \quad (7)$$

where $n$ and $k$ are the numbers of frames in the time and frequency domains, respectively. Furthermore, $T(n)$ is the time difference between two adjacent thermal images; $R(k)$ denotes the real part and $I(k)$ denotes the imaginary part. Thus, the frequency domain amplitude and phase of a single pixel can be obtained.

PCA transforms the original data into a set of linearly independent representations of each dimension through linear transformation, which can be used to extract the principal components of the data.

$$\hat{I} = U\Gamma V^T, \hat{I}(m\cdot n, N_t) = \frac{I(m\cdot n, N_t) - \mu_{m\cdot n}}{\sigma_{m\cdot n}} \quad (8)$$

$$\mu_{m\cdot n} = \frac{1}{N_t}\sum_{k=1}^{N_t} I(:,k)$$
$$\sigma_{m\cdot n}^2 = \frac{1}{N_t-1}\sum_{k=1}^{N_t}(I(:,k) - \mu_{m\cdot n})^2 \quad (9)$$

where $\Gamma$ is a singular value matrix of image data, $U$ and $V$ are unitary matrices, $I(m\cdot n, N_t)$ is a two-dimensional matrix converted from the image sequence, $m$ and $n$ are the numbers of rows and columns of images, respectively, and $N_t$ is the number of image frames. $\mu_{m\cdot n}$ and $\sigma_{m\cdot n}$ are the mean and standard deviation of the image matrix, respectively, and $k$ is the current frame number from 1 to $N_t$.

### III. SIMULATION

To investigate the influence of environmental thermal radiation intensity on the DoLP of the infrared radiation of the object's surface, a simulation was conducted based on (2). The DoLP of the infrared radiation of the object surface was simulated with different observation directions. Different types of targets were simulated by selecting different ranges for $\alpha$. The simulation conditions were set as follows: The incident angle of the environmental thermal radiation was 0°, i.e., the irradiation direction was directly facing the target. The refractive index and surface roughness of the different materials are listed in

TABLE I. The refractive index of air $n_{air}$ = 1.0 [30]. For an opaque surface, the complex refractive index is known.

TABLE I
REFRACTIVE INDEX AND SURFACE ROUGHNESS OF DIFFERENT MATERIALS [30],[31].

| | $n_m$ | $\sigma_m/(\mu m)$ |
|---|---|---|
| Aluminum | 4.88−1.17i | 0.63 |
| Nickel | 3.84−0.71i | 0.92 |
| Paper | 2.65 | 1.01 |
| CFRP | 2.77 | 1.11 |
| Rubber | 1.93 | 1.17 |

The target materials— aluminum, rubber, nickel, paper, and CFRP—were simulated and experimentally demonstrated. As shown in Fig. 3, to sufficiently demonstrate the influence of the material surface properties on the DoLP, we selected only the simulation results for aluminum, rubber, and CFRP for comparison. This is because the refractive indices and surface roughness values of these materials are significantly different, as shown in

TABLE I. Fig. 3 (a), (b), and (c) shows the infrared radiation DoLP of aluminum, rubber, and CFRP, respectively, under different $\alpha$ values. Notably, with an increase in the observation angle, the target infrared DoLP increased slowly at first, followed by a rapid decrease. In addition, the DoLP can reach a maximum value when the observation angle is approximately 80°. The objects were in a state of cooling after being subjected to short-term thermal excitations, and $\alpha$ was less than 1. When $\alpha$ is less than 1, the surface radiation DoLPs of the materials listed in

TABLE I, gradually decreases as $\alpha$ increases. This phenomenon indicates that the infrared polarization characteristics of the target object are significantly affected by $\alpha$. In addition, Fig. 3 shows that owing to the lower surface roughness and higher refractive index of aluminum, the DoLP of Aluminum is significantly greater than that of CFPR and rubber. The DoLP difference between rubber and CFPR was small and became less noticeable with an increase in $\alpha$ because the surface roughness values of both materials were similar.

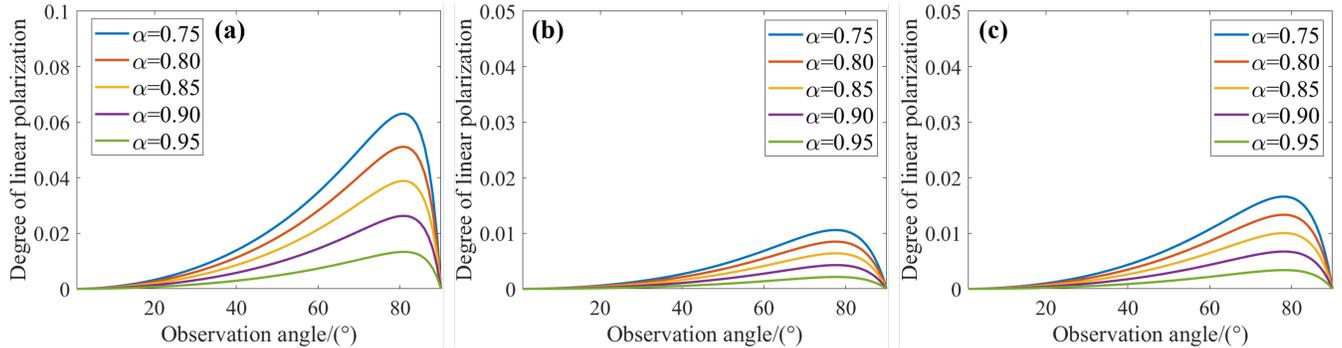

Fig. 3. Simulation results of the DoLP of surface infrared radiation for different materials: (a) aluminum, (b) rubber, and (c) CFRP.

The influence of the subsurface material of the target object on the infrared radiation polarization characteristics of the object surface is related to $\alpha$ [24]. When $\alpha$ is close to 1, the incident radiation intensity of the surrounding is approximately equal to that of the object. At this time, the polarization characteristics of the object's surface material play a major role, whereas the polarization characteristics of the substrate material under the surface play a minor role. Similarly, when $\alpha$ is close to 0, the polarization characteristics of the target material under the subsurface play a major role, whereas the polarization characteristics of the object surface material play a minor role. The weights of the infrared radiation polarization characteristics of the subsurface and surface are defined as $\omega_1$ and $\omega_2$, respectively. To simplify the mathematical simulation model, the relationship between $\omega_1$ and $\omega_2$ of the different materials in this study conforms to the relationship of the first-order function. Thus, the DoLP characteristic of the object surface is expressed as

$$DoLP_{simplified} = \omega_1 \times DoLP_{subsurface} + \omega_2 \times DoLP_{surface} \quad (10)$$

The simulation results are presented in Fig. 4. Notably, under the same $\alpha$, for the same surface material, the value and speed change of the DoLP on the aluminum material area were larger and faster than those on the rubber area. This implies that by measuring the polarization properties of infrared radiation in different regions of the object's surface, we can discern whether materials with different substrates or physical properties are present below the object's surface. NDT can be performed by measuring the change in the DoLP on the object surface. Furthermore, we can identify subsurface defects in different materials and distinguish the types of defects.

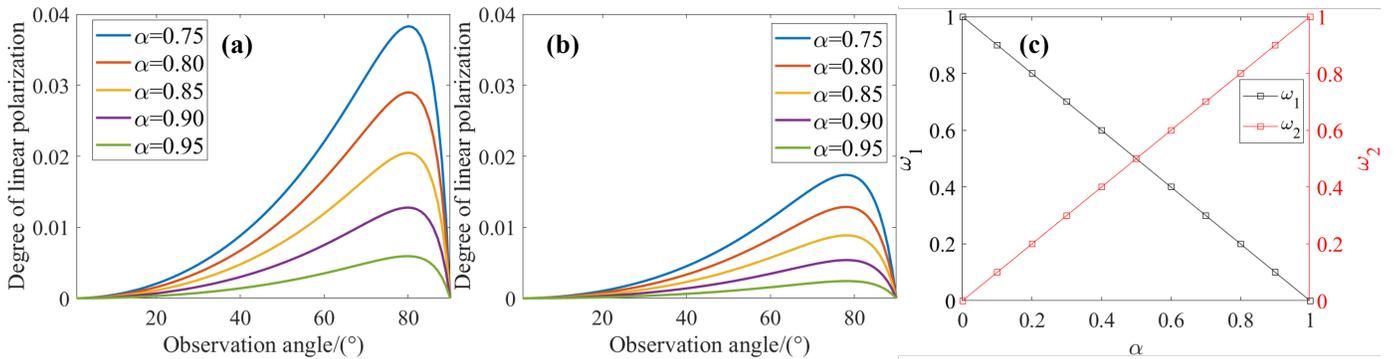

Fig. 4. Simulation results of surface DoLP when the subsurface artificial defect is made of different materials, the surface material is CFRP. Artificial defect simulated using (a) aluminum and (b) rubber. (c) Relationship between influence coefficients of material, $\omega_1$, $\omega_2$, and $\alpha$.

## IV. Experiments

### A. Experimental Setup

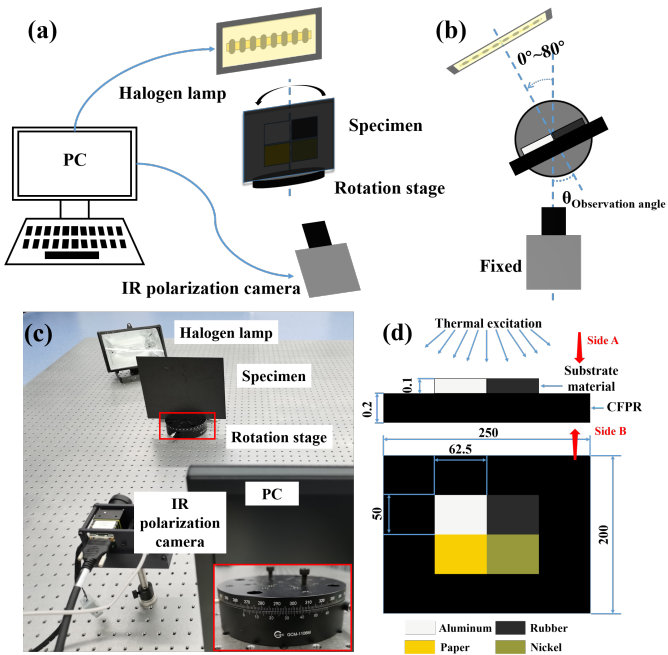

Fig. 5. Schematic and photos of experimental setup, (a) Schematic of infrared polarization camera and transmission NDT system. (b) Schematic of experimental setup (top view). The halogen lamp is pointed to side A of the specimen. (c) Photos of infrared polarization imaging-based transmission NDT system built in the laboratory. (d) Specimen used in the experiment.

An infrared polarization thermal imaging-based transmission NDT system was built for demonstration, as shown in Fig. 5. This system investigated the influence of surface materials and subsurface/substrate materials on the infrared polarization characteristics of the target object under a thermal radiation environment and verified the correctness of the simulation results. Subsequently, a series of pulsed infrared polarization thermal imaging NDT experiments was conducted. In this study, a CFRP specimen was created for experimental verification. Different materials were pasted onto the back of the specimen to simulate different types of subsurface artificial defects or substrates. A halogen lamp (1000 W) was used as the heat source to irradiate the back of the CFRP specimen for 1 s, as shown in Fig. 5 (d). The heat source can be considered a nonpolarized light source. The infrared polarization camera in the NDT system was an uncooled vanadium oxide infrared focal plane polarization detector (LD-LW640-P) produced by Xi'an Liding Optoelectronic Technology. The details of the infrared polarization camera are presented in Fig. 6 [32] and TABLE II. As shown in Fig. 6, the image captured by the DoFP polarization imaging camera is generally called a DoFP image. It not only contains the infrared radiation intensity information but also the polarization information of the object. The polarization direction of each pixel is different from that of its neighboring pixels. Generally, every four neighboring pixels covered with different polarizers comprise a 2×2 superpixel, and the polarization direction is clockwise through the following angles; 135°, 0°, 45°, and 90°.

In the experiment, the CFRP plate implemented for creating the specimen was made of a Toshiba T300 plain-weave carbon fiber plate, including one layer of carbon fiber lamina. The total thickness of the CFRP plate was 0.2 mm. Four types of materials (aluminum, rubber, nickel, and paper) were selected to simulate artificial defects. The size of each artificial defect is 62.5 × 50 × 0.1 mm3, as shown in Fig. 5 (d). The observation angle of the polarization camera significantly affected the polarization imaging results. To ensure that the transmission radiation intensity of the specimen is constant, as shown in Fig 5 (b), the relative positions of the halogen lamp and the specimen were fixed, so that the halogen lamp and the specimen surface were always parallel and equidistant, and then the image sequence of different observation angles under the same heat excitation can be obtained. To accurately control the observation angle of the polarization camera in the experiments, a manual rotation stage was used to adjust the direction of the specimen and obtain the imaging results of the CFRP specimen.

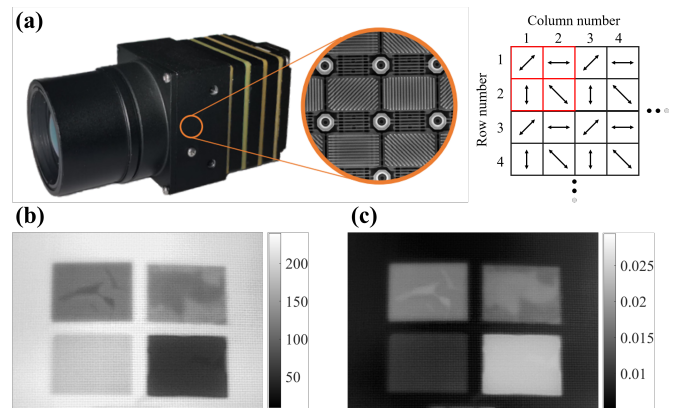

Fig. 6. (a) Schematic of polarization imaging principle of infrared polarization camera; (b) captured DoFP image, and (c) corresponding DoLP image.

TABLE II
PARAMETERS OF INFRARED POLARIZATION CAMERA

| Parameter | Value |
|---|---|
| Resolution of camera (pixel × pixel) | 640×512 |
| Wavelength range (nm) | 8000−14000 |
| Recording frequency (Hz) | 40 |
| Pixel size (μm) | 17 |
| focal length (mm) | 25 |

## B. Result

Two sets of experiments were conducted to verify the proposed method. By adjusting the angle of the rotating stage, an infrared polarization camera was used to measure the radiation polarization characteristics of sides A and B of the specimen at different angles after thermal excitation. The directions of sides A and B of the specimen are shown in Fig. 5 (d). The recorded infrared radiation polarization data of the specimens were used to verify the simulation results.

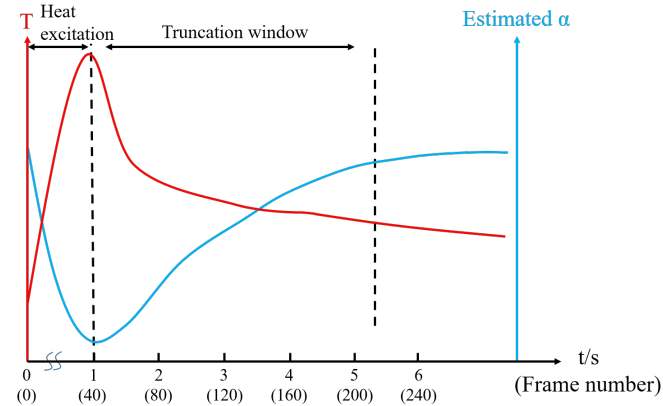

Fig. 7. Heating process and the truncation window of the experimental setup. The truncation window indicates the length of the thermogram sequences used for NDT data processing.

The heating process and truncation window used to select the image sequences for NDT data processing are shown in Fig. 7. All objects with a temperature above absolute zero (0 K) emit energy in the form of electromagnetic radiation. Therefore, during the heating process, the temperature of the object is higher than room temperature, and the corresponding $\alpha$ is always greater than 0 and less than 1. As the temperature of the object increases, the $\alpha$ value continues to decrease; however, it will not be less than 0. During the cooling process, because the testing specimen is subjected to environmental radiation, the reflected radiation still exists in Equations (1)–(4). As the temperature of the object decreases, the $\alpha$ value continues to increase close to but not exceeding 1 because the temperature of the object cannot be higher than the room temperature.

First, the DoLP simulation results obtained using (2) were verified experimentally. We selected 10 different angles at equal intervals between 0° and 90° and used a rotation stage to adjust the observation angle of the specimen to these specified angles. At each observation angle, a halogen lamp was used to heat side B of the specimen for 1 s, and an infrared polarization camera was used to record side A. The 40th and 80th frames of the polarization image sequence obtained after heating were used to calculate the corresponding DoLP images. Specified DoLP images captured at 10 different observation angles were generated, and the mean DoLP value in the region of interest was calculated, as shown in Fig. 8. Equation (2) is implemented for the curve fitting of the DoLP mean values shown in Fig. 8. Fig. 8 shows that, with an increase in the observation angle, the DoLP changes in different materials exhibit the same phenomenon. Within the range of approximately 0° to 80°, the DoLP increased slowly with an increase in the observation angle and reached a peak when the observation angle was approximately 80°. When an object is excited by an external heat source, the temperature of the target object begins to increase, alpha decreases, and the peak value of the DoLP curve increases accordingly. Simultaneously, the observation angle corresponding to the maximum value of the DoLP decreases. Conversely, when external thermal excitation is no longer applied to the object, the temperature of the target object begins to decrease, the value of $\alpha$ increases accordingly, the peak value of DoLP decreases, and the observation angle at which the peak value of DoLP appears gradually increases. The aforementioned experimental results verified the simulation results and were consistent with the phenomena described by the simulation results.

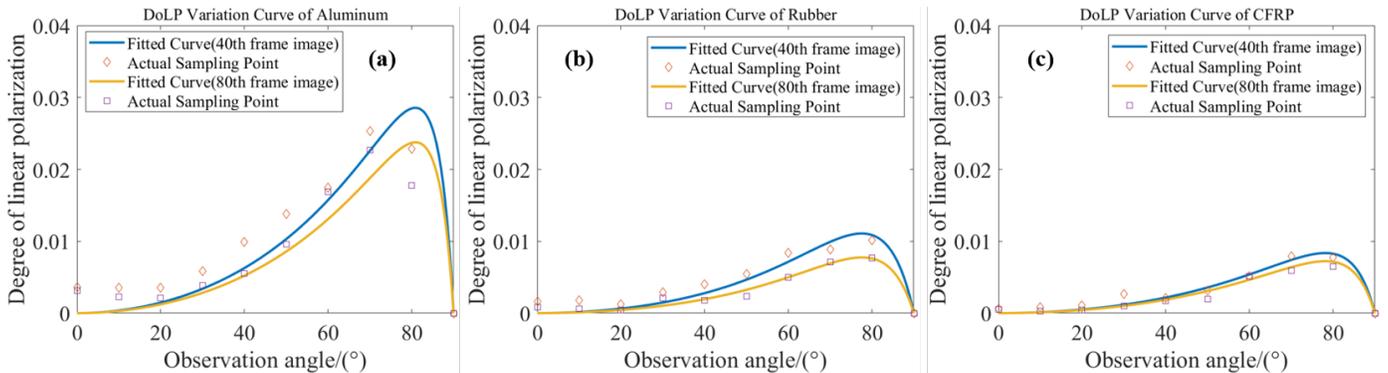

Fig. 8. Analysis of the 40th and 80th (corresponding time is 1s and 2s in Fig. 7) frames of the DoLP image sequence taken on side A of the specimen after heating starts. The mean DoLP values of (a) aluminum sheet, (b) rubber, and (c) CFRP in DoLP images captured at different angles was calculated, and the curves were fitted.

The simulation results using (10) were verified experimentally. The DoLP images were calculated using the image data of side B of the specimen captured using the infrared polarization camera. Curve fitting was performed using (10), as shown in Fig. 9. Herein, under the influence of subsurface material properties, such as those of aluminum and rubber, the change in the DoLP with respect to the observation angle also exhibits a phenomenon that first increases slowly and then decreases rapidly.

However, compared with the DoLP curve only under the influence of the CFRP material, as shown in Fig. 8(c), the surface DoLP of the defect area simulated by aluminum was numerically larger than that simulated by rubber.

The experimental results shown in Fig. 9 verify the hypothesis that the subsurface material affects the aforementioned polarization characteristics of the infrared radiation of the object surface, and the experimental results are consistent with the simulation results.

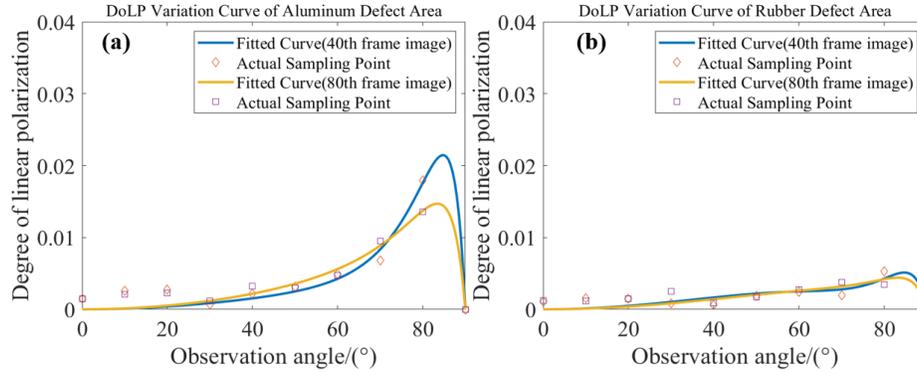

Fig. 9. Analysis of the 40th and 80th (corresponding time is 1s and 2s in Fig. 7) frames of the DoLP image sequence captured on side B of the specimen after heating. When the surface material is CFRP, different materials such as aluminum and rubber are used as sub-surface materials to simulate different types of defects. The DoLP changes on the surface of the specimen are recorded, and curve fitting is performed using (10). (a) The subsurface material is aluminum. (b) The subsurface material is rubber.

Subsequently, FFT and PCA processing were performed on the two images from side B of the specimen, DoFP intensity image and DoLP image sequences. The total number of frames was 175, which takes 4.375 s. The processing results are shown in Fig. 10 and Fig. 11. For the raw DoFP intensity image sequence and the DoLP image sequence recording the cooling process, we used only the first image in the sequence as a display. Notably, the edges of artificial defects on the DoLP images are clearer and sharper than those of artificial defects on intensity images, and the boundaries of defects are more accurate and significant. In the FFT processing results, we analyzed the phase images at $f = $ 0.23, 2.51, and 4.8 Hz. Notably, if the phase image obtained by the FFT method and the infrared radiation intensity image are identical, only a small difference exists between various materials at 0° and 10° when $f = $ 0.23 Hz, and the material characteristics cannot be distinguished. As $f$ increases, the difference between materials gradually increases. When $f = 4.8$ Hz, differences are observed between metal and non-metal at all angles. However, the difference between aluminum and nickel and that between paper and rubber cannot be further distinguished. Moreover, as $f$ increases, the overall noise of the image increases and the information at the edge of the artificial defects is weakened. For the phase images obtained based on the DoLP image sequence and FFT method, the FFT phase angle results maintain the polarization characteristics at $f = 0.23$ and 2.51 Hz. From 0° to 50°, the materials cannot easily be distinguished owing to their small and similar polarization values. However, from 60° to 80°, as the polarization degree of the materials increases and the difference between the DoLPs of the four artificial defect materials increases, the materials can be distinguished.

In the results obtained using the PCA method, because the defect features cannot be displayed well on the third principal component image and the subsequent principal component image, this study only displayed and analyzed the first and second principal components of the DoFP image sequence and DoLP image sequence, as shown in Fig. 10 and Fig. 11.

Compared with the detection result images obtained via the FFT method, the image obtained via the PCA method provides less noise and exhibits better anti-noise performance. Furthermore, the detection result image calculated using the DoLP image sequences shows a lower noise level than that calculated using the DoFP intensity image sequences. Moreover, the noise level of the detection result image obtained by the PCA method and the DoLP image sequences have the lowest noise level, indicating that the proposed DoLP image sequence-based NDT method provides excellent anti-noise performance.

Fig. 10 and Fig. 11 show that the detection results obtained using the FFT method and DoFP intensity image sequence cannot easily distinguish the four different artificially simulated defect areas and the noise level is high. In the detection results based on the DoLP image sequence and FFT method, when the observation angle exceeds 60 °, the NDT results showed a good ability to distinguish different types of defects. By analyzing the last row of the images in Fig. 11, that is, the PCA second principal component image, when the observation angle exceeds 60°, the differences between the four artificially simulated defect areas become significant. Although the sizes and thicknesses of the artificial defects are the same, in the detection results based on the DoLP image sequence, significant numerical differences, which can be distinguished intuitively, are observed between the defect areas. This is more conducive to examining these areas.

To demonstrate the advantages of the proposed NDT method, an infrared polarization information-based NDT method, the detection result images at an observation angle of 70° are shown in Fig. 12. Notably, the proposed method provides more accurate defect position and edge information than an infrared intensity image-based NDT method and avoids the influence of thermal diffusion to a certain extent. In particular, in the phase image obtained by combining DoLP images with the FFT method, different materials of subsurface artificial defects can be further distinguished in the $f = 2.51$Hz phase diagram. In addition, when the frequency of the phase image was low, the deep defects of the specimen were mainly detected; thus, the difference in the detection results of different artificial defective areas was high. When the frequency of the phase image was high, it mainly reflected the detection results of shallow defects, and the similarity of detection results of artificial defects with different material was high.

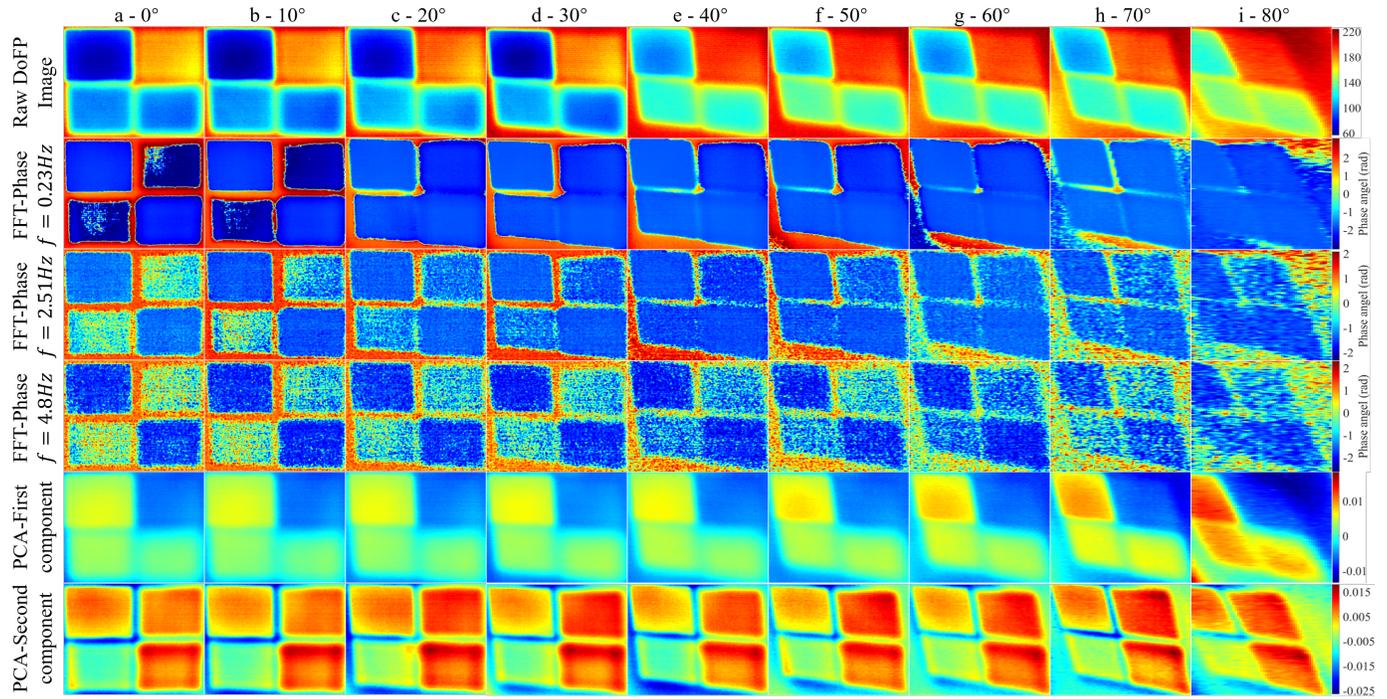

Fig. 10. NDT results based on infrared DoFP image sequence from B side of specimen. The first row is the first frame of the original intensity image sequence captured by the infrared polarization camera during the cooling process. Rows 2-4 are the phase images generated using FFT at $f = $ 0.23, 2.51, and 4.8 Hz; Rows 5-6 are the first and second principal component images calculated using PCA methods for the raw infrared polarization camera image sequence. From column (a) to (i), each column is the detection result of different observation angles, from 0° to 80°.

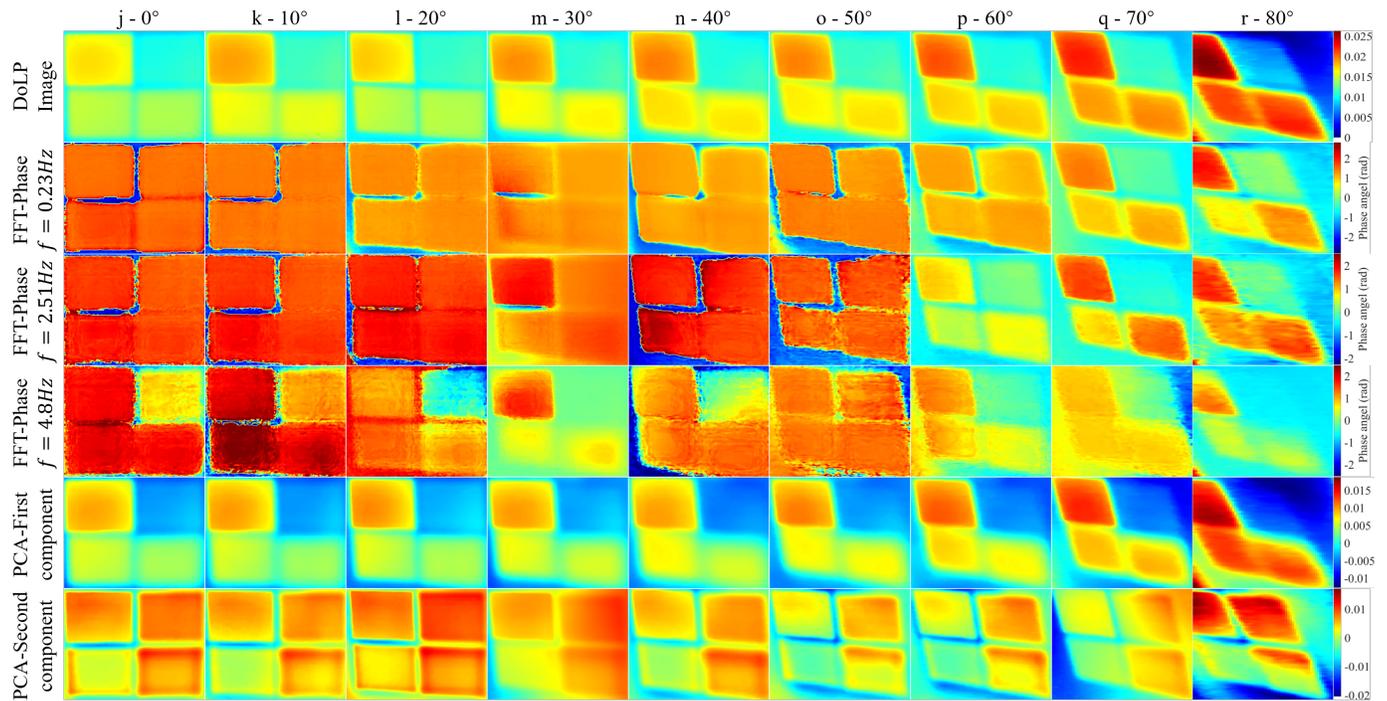

Fig. 11. NDT results generated using the DoLP image sequence from B side of specimen. The first row is the first frame of the DoLP image sequence captured by the infrared polarization camera during the cooling process. Rows 2-4 are the phase images at $f = $ 0.23, 2.51, and 4.8 Hz calculated using FFT for the DoLP image sequence; Rows 5-6 are the first and second principal component images of the DoLP image sequence calculated using the PCA method. From column (j) to (r), each column is the detection result of different observation angles, from 0° to 80°.

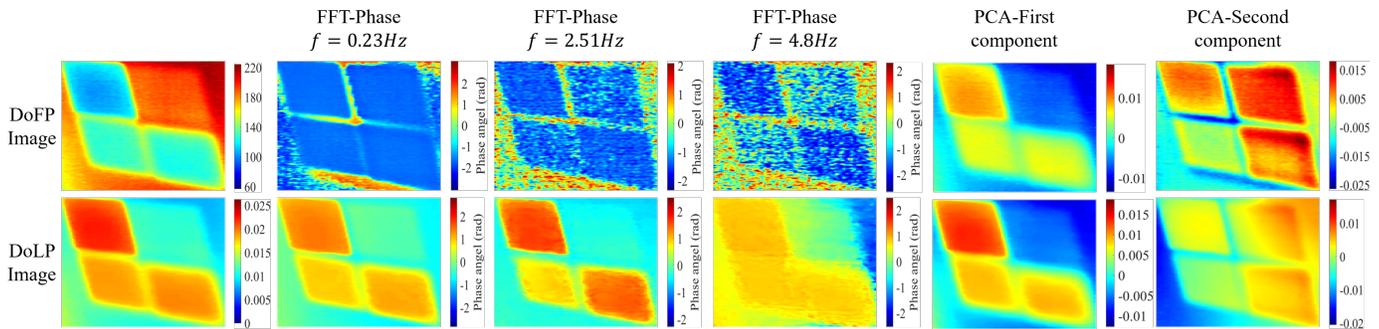

Fig. 12. Comparison of the NDT results of the two methods, when the observation angle is 70° (the first row is selected from column h of Fig. 10, and the second row is from column q of Fig. 11)

## V. CONCLUSION

In this study, an infrared polarization imaging technique based on an NDT method was proposed and demonstrated.

We demonstrated that the subsurface material and defects of an object affected the polarization characteristics of the object's surface radiation. The simulation and experimental results verified that subsurface features influenced the DoLP of the object's surface infrared radiation. Therefore, NDT inspection could be performed according to the changes in the DoLP of objects. When the temperature difference between the defective and non-defective areas was small, the proposed method could image the defective area.

Subsequently, a CFRP specimen was designed and built for the experiments. Four different types of subsurface defects, created using four different materials, were designed and embedded in the CFRP specimen for the experiments. The experimental results showed that the changes in the DoLP of the target defects and nondefective areas were consistent with the simulation results. Pulsed thermography NDT experiments were conducted for demonstration purposes. The thermogram and DoLP image sequences of the specimens were calculated and processed using FFT and PCA. The results showed that the proposed infrared polarization imaging-based NDT method is superior to conventional thermography-based NDT methods in terms of defect edge texture, and detection of the defect position, particularly at specific observation angles where the DoLP changes rapidly. In particular, from the FFT results obtained at specific observation angles, the proposed method showed the potential to distinguish subsurface artificial defect materials. In this study, we demonstrated the application prospects of infrared polarization imaging techniques in the field of NDT and provided a new method for further research on infrared thermography NDT methods.

## REFERENCE

[1] V. P. Vavilov and D. D. Burleigh, "Review of pulsed thermal NDT: Physical principles, theory and data processing," *Ndt & E International,* vol. 73, pp. 28-52, 2015.

[2] D. Bates, G. Smith, D. Lu, and J. Hewitt, "Rapid thermal non-destructive testing of aircraft components," *Composites Part B: Engineering,* vol. 31, no. 3, pp. 175-185, 2000.

[3] X. Wu and K. Peters, "Non-destructive inspection of adhesively bonded joints using amplitude modulated thermography," *Experimental Mechanics,* vol. 55, no. 8, pp. 1485-1501, 2015.

[4] N. Puthiyaveettil, K. R. Thomas, S. Unnikrishnakurup, P. Myrach, M. Ziegler, and K. Balasubramaniam, "Laser line scanning thermography for surface breaking crack detection: Modeling and experimental study," *Infrared Physics & Technology,* vol. 104, p. 103141, 2020.

[5] J.-H. He, D.-P. Liu, C.-H. Chung, and H.-H. Huang, "Infrared Thermography Measurement for Vibration-Based Structural Health Monitoring in Low-Visibility Harsh Environments," *Sensors,* vol. 20, no. 24, p. 7067, 2020.

[6] Z. Liu, B. Gao, and G. Y. Tian, "Natural Crack Diagnosis System Based on Novel L-Shaped Electromagnetic Sensing Thermography," *IEEE Transactions on Industrial Electronics,* vol. 67, no. 11, pp. 9703-9714, 2020.

[7] B. Gao, L. Bai, W. L. Woo, G. Y. Tian, and Y. Cheng, "Automatic Defect Identification of Eddy Current Pulsed Thermography Using Single Channel Blind Source Separation," *IEEE Transactions on Instrumentation and Measurement,* vol. 63, no. 4, pp. 913-922, 2014.

[8] V. P. Vavilov, D. D. J. N. Burleigh, and E. International, "Review of pulsed thermal NDT: Physical principles, theory and data processing," vol. 73, pp. 28-52, 2015.

[9] L. Yu, Y. Guo, H. Zhu, M. Luo, and X. J. M. Ji, "Low-Cost Microbolometer Type Infrared Detectors," vol. 11, no. 9, p. 800, 2020.

[10] T. J. Rogne, F. G. Smith, and J. E. Rice, "Passive target detection using polarized components of infrared signatures," in *Polarimetry: Radar, Infrared, Visible, Ultraviolet, and X-Ray*, 1990, vol. 1317, pp. 242-251: SPIE.

[11] H. JIANG, F. Qiang, and J. DUAN, "Research on infrared polarization imaging detection technologyand application," *Infrared Technology,* vol. 36, no. 5, pp. 345-349, 2014.

[12] R. M. Matchko and G. R. Gerhart, "High-speed imaging chopper polarimetry," *Optical Engineering,* vol. 47, no. 1, p. 016001, 2008.

[13] L. Giudicotti and M. Brombin, "Data analysis for a rotating quarter-wave, far-infrared Stokes polarimeter," *Applied optics,* vol. 46, no. 14, pp. 2638-2648, 2007.

[14] J. S. Tyo, D. L. Goldstein, D. B. Chenault, and J. A. Shaw, "Review of passive imaging polarimetry for remote sensing applications," *Applied optics,* vol. 45, no. 22, pp. 5453-5469, 2006.

[15] M. Kimata, "Uncooled infrared focal plane arrays," *IEEJ Transactions on Electrical and Electronic Engineering,* vol. 13, no. 1, pp. 4-12, 2018.

[16] W. Xia, X. Runqiu, J. Weiqi, L. Jing, and L. Jian'an, "Technology progress of infrared polarization imaging detection," *Infrared and Laser Engineering,* vol. 43, no. 10, pp. 3175-3182, 2014.

[17] C. Wei-Li, W. Shu-Hua, J. Wei-Qi, and L. Jun-Wei, "Research of infrared polarization characteristics based on polarization micro-surface theory," *Journal of Infrared and Millimeter Waves,* vol. 33, no. 5, pp. 507-514, 2014.

[18] A. G. Andreou and Z. K. Kalayjian, "Polarization imaging: principles and integrated polarimeters," *IEEE Sensors journal,* vol. 2, no. 6, pp. 566-576, 2002.

[19] Kristan, Gurton, Rachid, and D. J. A. Optics, "Effect of surface roughness and complex indices of refraction on polarized thermal emission," 2005.

[20] A. Hs *et al.*, "Analysis of infrared polarization properties of targets with rough surfaces - ScienceDirect," vol. 151.

[21] L. B. Wolff, A. Lundberg, and R. Tang, "Image understanding from thermal emission polarization," in *Proceedings. 1998 IEEE Computer Society Conference on Computer Vision and Pattern Recognition (Cat. No. 98CB36231)*, 1998, pp. 625-631: IEEE.

[22] F. Liu, X. Shao, Y. Gao, B. Xiangli, P. Han, and G. Li, "Polarization characteristics of objects in long-wave infrared range," (in eng), *Journal of the Optical Society of America. A, Optics, image science, and vision,* vol. 33, no. 2, pp. 237-243, 2016/02// 2016.

[23] L. Liu, L. Liu, R. Ming, S. Wu, and J. Zhang, "An infrared DoLP model considering the radiation coupling effect," in *Photonics*, 2021, vol. 8, no. 12, p. 546: MDPI.

[24] X. Wenbin *et al.*, "Research on coating materials detection and recognition based on infrared spectral polarization degree contrast," - *Infrared and Laser Engineering,* vol. - 49, no. - 6, pp. - 20190445-1, - 2020-07-01 2020.

[25] H. Shi, Y. Liu, C. He, C. Wang, Y. Li, and Y. Zhang, "Analysis of infrared polarization properties of targets with rough surfaces," *Optics & Laser Technology,* vol. 151, p. 108069, 2022/07/01/ 2022.

[26] X. Maldague and S. J. J. o. a. p. Marinetti, "Pulse phase infrared thermography," vol. 79, no. 5, pp. 2694-2698, 1996.

[27] N. Rajic, "Principal component thermography for flaw contrast enhancement and flaw depth characterisation in composite structures," *Composite Structures,* vol. 58, no. 4, pp. 521-528, 2002/12/01/ 2002.

[28] S. Gao and V. Gruev, "Bilinear and bicubic interpolation methods for division of focal plane polarimeters," *Optics Express,* vol. 19, no. 27, pp. 26161-26173, 2011/12/19 2011.

[29] S. Liu, J. Chen, Y. Xun, X. Zhao, and C. H. Chang, "A New Polarization Image Demosaicking Algorithm by Exploiting Inter-Channel Correlations With Guided Filtering," *IEEE Transactions on Image Processing,* vol. 29, pp. 7076-7089, 2020.

[30] J. Huilin, L. Yi, and S. Haodong, "Infrared polarization properties of targets with rough surface," *Chinese Optics,* vol. 13, no. 3, pp. 459-471, 2020.

[31] J. H. Weaver and H. P. R. Frederikse, "Optical properties of selected elements," *CRC Handbook of Chemistry and Physics,* pp. 12-133, 01/01 2001.

[32] N. Li, Y. Zhao, R. Wu, and Q. Pan, "Polarization-guided road detection network for LWIR division-of-focal-plane camera," *Optics Letters,* vol. 46, no. 22, pp. 5679-5682, 2021/11/15 2021.